\documentclass[prl,twocolumn,showpacs,aps]{revtex4}
\usepackage{amsmath}
\usepackage{amssymb}
\usepackage[dvips]{graphicx}

\begin{document}

\title{Spin dynamics in electron-doped pnictide superconductors}

\author{Yi Gao$^{1}$, Tao Zhou$^{1,2}$, C. S. Ting$^{1,3}$, and Wu-Pei Su$^{1}$}
\affiliation{$^{1}$Department of Physics and Texas Center for
Superconductivity, University of Houston, Houston, Texas, 77204,
USA\\
$^{2}$Department of Physics, Nanjing University of Aeronautics and
Astronautics, Nanjing, 210016, China\\
$^{3}$Department of Physics, Fudan University, Shanghai, 200433,
China}

\begin{abstract}
The doping dependence of spin excitations in
Ba(Fe$_{1-x}$Co$_{x}$)$_{2}$As$_{2}$ is studied based on a
two-orbital model under RPA approximation. The interplay between the
spin-density-wave (SDW) and superconductivity (SC) is considered in
our calculation. Our results for the spin susceptibility are in good
agreement with neutron scattering (NS) experiments in various doping
ranges at temperatures (T) above and below the superconducting
transition temperature T$_{c}$. For the overdoped sample where one
of the two hole pockets around $\Gamma$ point disappears according
to ARPES, we show that the imaginary part of the spin susceptibility
in both SC and normal phases exhibits a gap-like behavior. This
feature is consistent with the ``pseudogap'' as observed by recent
NMR and NS experiments.

\end{abstract}


\maketitle

The recent discovery of the iron arsenide superconductors \cite{I1},
whose parent compounds exhibit long-range antiferromagnetic (AF) or
spin-density-wave (SDW) order similar to the cuprates \cite{I2},
provides another promising group of materials for studying the
interplay between magnetism and superconductivity (SC). Especially,
the electron-doped pnictide superconductors like
Ba(Fe$_{1-x}$Co$_{x}$)$_{2}$As$_{2}$ \cite{I3} has emerged as one of
the most important systems due to the availability of large
homogeneous single crystals. The phase diagram \cite{I4,I5,I6-1} for
these materials indicates that the parent compound upon cooling
through T$_{N}\sim$ 140K \cite{I7} develops a static SDW order.
Increasing the doping of Co, the SDW order is suppressed and the SC
order emerges as the temperature (T) falls below T$_{c}$. The SDW
and SC orders coexist in the underdoped samples
\cite{I4,I5,I6-1,I10}. By further increasing the Co concentration to
the optimally doped regime, the SDW order disappears. These
experimental results provide compelling evidences for strong
competition between the SDW and SC orders.

Recently, several neutron scattering (NS) experiments have been
carried out to probe the spin dynamics in these materials
\cite{I4,I6-1,I10,I12,I13,I14,I15,I16,I17,I18}, and the spin
excitation spectrum was fitted by using a $J_{1}, J_{2}$ Heisenberg
model based on localized spins \cite{I12,I10,I18}. However, while
the parent compounds of the cuprates are Mott insulators with large
in-plane exchange interactions \cite{I19}, the parent iron arsenides
are bad metals and remain itinerant at all doping levels. Magnetism
in these materials are most likely to originate from itinerant
electrons and the AF order is a result of SDW instability due to
Fermi-surface nesting \cite{I20}. Theoretically, at present, the
variation of the spin susceptibility with doping remains less
explored. The spin susceptibilities were mostly studied in the
optimally doped compounds without SDW \cite{I21} or in the parent
compound without SC \cite{I21-1}. In this work, we adopt
Fermi-liquid mean field (MF) theory to study the static SDW and SC,
and employ the random-phase approximation (RPA) to investigate the
spin dynamics in Ba(Fe$_{1-x}$Co$_{x}$)$_{2}$As$_{2}$  from  the
imaginary part of the dynamic spin susceptibility. It is expected
that the present approach is much more justified than the localized
model for examining the spin fluctuations in the iron arsenides, as
suggested by both experiments \cite{I13,I16} and theories
\cite{I21,I21-1}. We show that the calculated spin susceptibilities
are in qualitative agreement with several NS and NMR experiments in
various doping ranges.

We start with a two-orbital model by taking into account two Fe ions
per unit cell \cite{M1}. The reason we adopt this model is its
ability \cite{M4} to qualitatively account for the doping evolution
of the Fermi surface and the asymmetry in the SC coherent peaks as
observed by the angle resolved photo-emission spectroscopy (ARPES)
\cite{M2} and the scanning tunneling microscopy (STM) \cite{M3}
experiments on Ba(Fe$_{1-x}$Co$_{x}$)$_{2}$As$_{2}$. The Hamiltonian
of our system can be expressed as \cite{M4}
$H=H_{0}+H_{\Delta}+H_{int}$. $H_{0}$ is the tight-binding
Hamiltonian and can be written as
$H_{0}=\sum_{\mathbf{k}\sigma}\psi_{\mathbf{k}\sigma}^{\dag}M_{\mathbf{k}}\psi_{\mathbf{k}\sigma}$
\cite{M1,M4}, where
$\psi_{\mathbf{k}\sigma}^{\dag}=(c_{A0,\mathbf{k}\sigma}^{\dag},c_{A1,\mathbf{k}\sigma}^{\dag},
c_{B0,\mathbf{k}\sigma}^{\dag},c_{B1,\mathbf{k}\sigma}^{\dag})$ is
the creation operator with spin $\sigma=-1$ ($\downarrow$) or $1$
($\uparrow$), in the orbitals $(0,1)=(d_{xz},d_{yz})$ at the
sublattice $A$ ($B$), and
\begin{equation}
\label{c} M_{\mathbf{k}}=\begin{pmatrix}
\varepsilon_{A,\mathbf{k}}-\mu&\varepsilon_{xy,\mathbf{k}}&\varepsilon_{T,\mathbf{k}}&0\\
\varepsilon_{xy,\mathbf{k}}&\varepsilon_{A,\mathbf{k}}-\mu&0&\varepsilon_{T,\mathbf{k}}\\
\varepsilon_{T,\mathbf{k}}&0&\varepsilon_{B,\mathbf{k}}-\mu&\varepsilon_{xy,\mathbf{k}}\\
0&\varepsilon_{T,\mathbf{k}}&\varepsilon_{xy,\mathbf{k}}&\varepsilon_{B,\mathbf{k}}-\mu
\end{pmatrix},
\end{equation}
where $\varepsilon_{A,\mathbf{k}}=-2(t_{2}\cos k_{x}+t_{3}\cos
k_{y})$, $\varepsilon_{B,\mathbf{k}}=-2(t_{2}\cos k_{y}+t_{3}\cos
k_{x})$, $\varepsilon_{xy,\mathbf{k}}=-2t_{4}(\cos k_{x}+\cos
k_{x})$, and $\varepsilon_{T,\mathbf{k}}=-4t_{1}\cos
\frac{k_{x}}{2}\cos \frac{k_{y}}{2}$. $t_{1-4}$ are the hopping
parameters and $\mu$ is the chemical potential. Throughout the
paper, the momentum $\mathbf{k}$ is defined in the tetragonal
notation.

The pairing term is
$H_{\Delta}=\sum_{\mathbf{k}s\alpha}(\Delta_{\mathbf{k}}
c_{s\alpha,\mathbf{k}\uparrow}^{\dag}c_{s\alpha,-\mathbf{k}\downarrow}^{\dag}+h.c.)$.
Here, we assume there exits only next-nearest-neighbor intraorbital
pairing with extended $s-$wave pairing symmetry, and the SC order
parameter is $\Delta_{\mathbf{k}}=\frac{\Delta_{0}}{2}(\cos
k_{x}+\cos k_{y})$ similar to that obtained by spin fluctuations
\cite{I20}, where
\begin{equation}
\label{f} \Delta_{0}=\frac{2V_{nnn}}{N}\sum_{\mathbf{k}}(\cos
k_{x}+\cos k_{y})\langle
c_{s\alpha,-\mathbf{k}\downarrow}c_{s\alpha,\mathbf{k}\uparrow}\rangle,
\end{equation}
with $V_{nnn}$ being the attractive pairing interaction.

$H_{int}$ is the on-site interaction term which includes the
Coulombic interaction and Hund coupling $J_{H}$, following Refs.
\cite{M4,M5,I21}, it can be expressed as
\begin{eqnarray}
\label{d}
H_{int}=&U&\sum_{ijs\alpha}n_{s\alpha,ij\uparrow}n_{s\alpha,ij\downarrow}
+(U^{'}-\frac{J_{H}}{2})\sum_{ijs}n_{s0,ij}n_{s1,ij}\nonumber\\
&+&J_{H}\sum_{ijs}(c_{s0,ij\uparrow}^{\dag}c_{s0,ij\downarrow}^{\dag}c_{s1,ij\downarrow}c_{s1,ij\uparrow}+h.c.)\nonumber\\
&-&2J_{H}\sum_{ijs}\mathbf{S}_{s0,ij}\cdot\mathbf{S}_{s1,ij},
\end{eqnarray}
where $\{i,j\}$ denotes the unit cell, $s=0$ ($A$) or $1$ ($B$) is
the sublattice index, and $\alpha=0$ ($d_{xz}$) or $1$ ($d_{yz}$)
represents the orbital. $n_{s\alpha,ij\sigma}$ and
$\mathbf{S}_{s\alpha,ij}$ are the density and spin operators in the
orbital $\alpha$ at the sublattice $s$ of the unit cell $\{i,j\}$,
respectively. According to symmetry, we have $U^{'}=U-2J_{H}$
\cite{M5}. In the MF approach, we linearize $H_{int}$ in momentum
space as \cite{M4,M6}
\begin{eqnarray}
\label{d2}
H_{int}^{MF}&=&\frac{n}{4}(3U-5J_{H})\sum_{\mathbf{k}s\alpha\sigma}n_{s\alpha,\mathbf{k}\sigma}\nonumber\\
&-&\frac{m}{2}(U+J_{H})\sum_{\mathbf{k}s\alpha\sigma}\sigma
c_{s\alpha,\mathbf{k}+\mathbf{Q}\sigma}^{\dag}c_{s\alpha,\mathbf{k}\sigma},
\end{eqnarray}
where $n=2+x$ is the number of electrons per lattice site and
$\mathbf{Q}=(\pi,\pi)$. The SDW order parameter is
$m=\frac{1}{N}\sum_{\mathbf{k}\sigma}\sigma\langle
c_{s\alpha,\mathbf{k}+\mathbf{Q}\sigma}^{\dag}c_{s\alpha,\mathbf{k}\sigma}\rangle$,
with $N$ being the number of unit cells.

The effective MF Hamiltonian is then given by
\begin{eqnarray}
\label{g}
H^{MF}&=&\sum_{\mathbf{k}}\hspace{0.2mm}'\varphi_{\mathbf{k}}^{\dag}W_{\mathbf{k}}\varphi_{\mathbf{k}},\nonumber\\
W_{\mathbf{k}}&=&\begin{pmatrix}
M_{\mathbf{k}}^{'}&R&\Delta_{\mathbf{k}}I&0\\
R&M_{\mathbf{k}+\mathbf{Q}}^{'}&0&-\Delta_{\mathbf{k}}I\\
\Delta_{\mathbf{k}}I&0&-M_{\mathbf{k}}^{'}&R\\
0&-\Delta_{\mathbf{k}}I&R&-M_{\mathbf{k}+\mathbf{Q}}^{'}
\end{pmatrix},
\end{eqnarray}
where
$\varphi_{\mathbf{k}}^{\dag}=(\psi_{\mathbf{k}\uparrow}^{\dag},\psi_{\mathbf{k}+\mathbf{Q}\uparrow}^{\dag},
\psi_{-\mathbf{k}\downarrow},\psi_{-(\mathbf{k}+\mathbf{Q})\downarrow})$,
$M_{\mathbf{k}}^{'}=M_{\mathbf{k}}+\frac{n}{4}(3U-5J_{H})I$, and
$R=-\frac{m}{2}(U+J_{H})I$. $I$ is a $4\times4$ unit matrix and
$\sum_{\mathbf{k}}^{'}$ means the summation extends over the
magnetic Brillouin zone (MBZ): $-\pi<k_{x}\pm k_{y}\leqslant\pi$.
The MF Green's function matrix can be written as
$g(\mathbf{k},\tau)=-\langle
T_{\tau}\varphi_{\mathbf{k}}(\tau)\varphi_{\mathbf{k}}^{\dag}(0)\rangle$
and
$g(\mathbf{k},ip_{n})=A_{\mathbf{k}}W_{\mathbf{k}}^{'}A_{\mathbf{k}}^{\dag}$,
where
$W_{\mathbf{k}ij}^{'}=\delta_{ij}(ip_{n}-\lambda_{\mathbf{k}i})^{-1}$
and $A_{\mathbf{k}}$ is a unitary matrix that satisfies
$(A_{\mathbf{k}}^{\dag}W_{\mathbf{k}}A_{\mathbf{k}})_{ij}=\delta_{ij}\lambda_{\mathbf{k}i}$.

\begin{figure}
\includegraphics[width=1\linewidth]{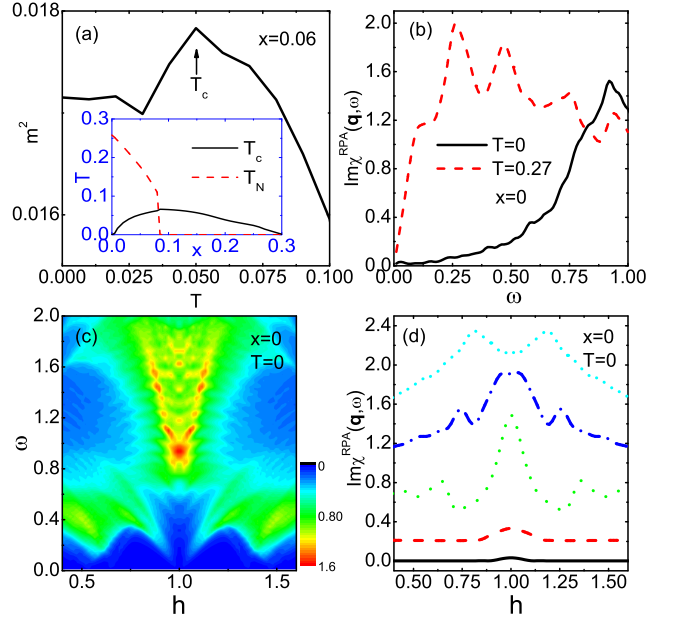}
\caption{\label{phase diagram} (Color online) (a) $m^{2}$ as a
function of $T$ close to $T_{c}$ at $x=0.06$. Inset shows the
calculated phase diagram. (b) $Im\chi^{RPA}(\mathbf{q},\omega)$ at
$\mathbf{q}=(\pi,\pi)$ as a function of $\omega$ by changing
$i\omega_{n}$ to $\omega+i\eta$, at $x=0$ and different temperatures
$T$. (c) $Im\chi^{RPA}(\mathbf{q},\omega)$ as a function of energy
transfer $\omega$ and momentum $\mathbf{q}$, at $x=0$ and $T=0$. The
momentum is scanned along $(q_{x}/\pi,q_{y}/\pi)=(h,h)$. (d)
Constant-energy scans along the $(q_{x}/\pi,q_{y}/\pi)=(h,h)$
direction at $x=0$ and $T=0$. Successive cuts are displaced
vertically for clarity. The energy transfer is $\omega=0.09$ (black
solid), $0.4$ (red dash), $0.93$ (green dot), $1.2$ (blue dash dot),
and $2$ (cyan short dash). The damping rate $\eta=0.04$.}
\end{figure}

First, we solve the MF equations self-consistently to obtain $m$,
$\Delta_{0}$ and $\mu$ at different doping levels $x$ and
temperatures $T$. The magnitudes of the parameters are chosen as
$t_{1-4}=1,0.4,-2,0.04$ \cite{M1}, $U=3.4$, $J_{H}=1.3$,
$V_{nnn}=-1.2$ \cite{M4}, and the number of unit cells is
$257\times257$. Throughout the paper, the energies are measured in
units of $|t_{1}|$. The calculated phase diagram as shown in the
inset of Fig. \ref{phase diagram}(a) reproduces the result based on
Bogoliubov-de Gennes (BdG) equations \cite{M4} and is also
consistent with the experiments on
Ba(Fe$_{1-x}$Co$_{x}$)$_{2}$As$_{2}$ \cite{I4,I5,I6-1}. Here the SDW
and SC are competing with each other. If there is no SDW, SC would
show up even in the parent compound. The presence of SC also
suppresses SDW. For example, in the underdoped ($x=0.06$) compound
with $T_{N}\approx0.16$ and $T_{c}\approx0.05$, the calculated
magnitude of $m^{2}$ (proportional to the magnetic Bragg peak
intensity) at $T=0$ is reduced by $\sim4\%$ relative to that of the
maximum intensity at $T_{c}$ [see Fig. \ref{phase diagram}(a)], and
this result is consistent with the neutron diffraction experiments
\cite{I4,I10}.

Then, we investigate the spin dynamics in
Ba(Fe$_{1-x}$Co$_{x}$)$_{2}$As$_{2}$ for $x=0$, $0.06$, $0.1$, and
$0.2$, corresponding to the undoped, underdoped, optimally doped and
overdoped compounds, respectively. The MF spin susceptibility is
$\chi^{r\alpha,s\beta(0)}_{t\gamma,u\delta}(\mathbf{q},\mathbf{q}^{'},i\omega_{n})=
\frac{\delta_{\mathbf{q}^{'},\mathbf{q}}}{2}\sum_{(i,j)=(p,o)}^{(o+8,p+8)}
[P_{im,nj}(q)+P_{i+4m,nj+4}(q)]$, here, $r,s,t,u$ label the
sublattice indices, $\alpha,\beta,\gamma,\delta$ represent the
orbitals, $m=2r+\alpha+1,n=2s+\beta+1,o=2t+\gamma+1,p=2u+\delta+1$,
and $P_{im,nj}(q)=-\frac{1}{\beta
N}\sum_{\mathbf{k},p_{n}}g_{im}(k)g_{nj}(k+q)$. Here we used
$k=(\mathbf{k},ip_{n})$ and $q=(\mathbf{q},i\omega_{n})$.

We then use RPA to take into account the residual fluctuation of
$H_{int}$ beyond MF. The RPA spin susceptibility is determined by
the matrix equation $\chi^{RPA}(q)=\sum_{rt\alpha\gamma}\{
\chi^{0}(q)[I-\Gamma\chi^{0}(q)]^{-1}
\}^{r\alpha,r\alpha}_{t\gamma,t\gamma}$, where $I$ is a $16\times16$
unit matrix and the nonzero elements of the interaction vertex are:
for $\alpha=\beta=\gamma=\delta$,
$\Gamma^{r\alpha,r\beta}_{r\gamma,r\delta}=2U$; for
$\alpha=\beta\neq\gamma=\delta$ or $\alpha=\gamma\neq\beta=\delta$,
$\Gamma^{r\alpha,r\beta}_{r\gamma,r\delta}=2J_{H}$.

\begin{figure}
\includegraphics[width=1\linewidth]{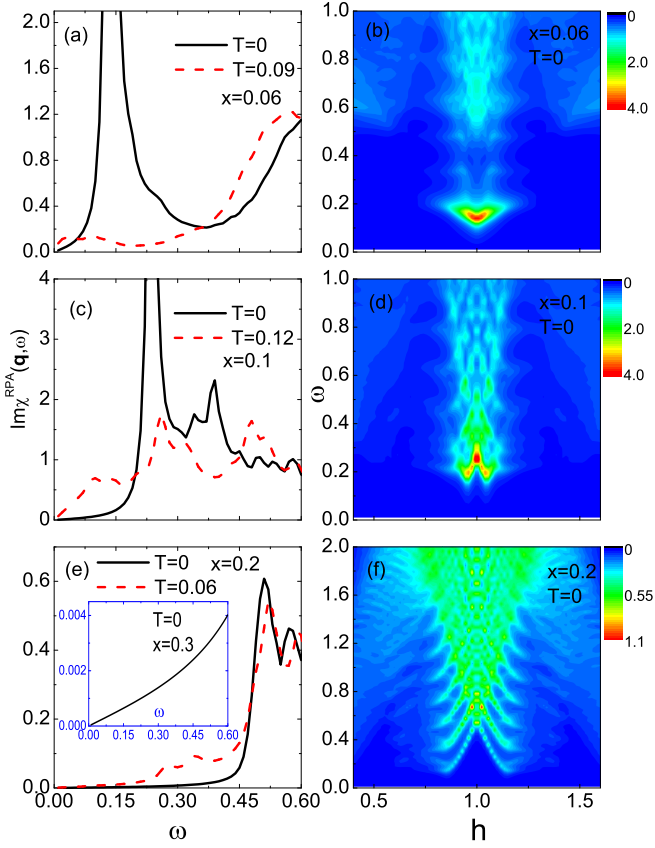}
 \caption{\label{Constant-Q scan} (Color online)
(a) $Im\chi^{RPA}(\mathbf{q},\omega)$ at $\mathbf{q}=(\pi,\pi)$ as a
function of $\omega$, at $x=0.06$ and different temperatures $T$.
(b) $Im\chi^{RPA}(\mathbf{q},\omega)$ as a function of energy
transfer $\omega$ and momentum $\mathbf{q}$, at $x=0.06$ and $T=0$.
The momentum is scanned along $(q_{x}/\pi,q_{y}/\pi)=(h,h)$. (c) and
(d) ((e) and (f)) are similar to (a) and (b), respectively, but at
$x=0.1$ ($x=0.2$). Inset in (e) shows the $x=0.3$ case.}
\end{figure}

In the parent ($x=0$) compound, the RPA spin susceptibility [see
Fig. \ref{phase diagram}(b)] in the paramagnetic state at $T=0.27$
($T_{N}\approx0.25$) shows a linear energy dependence for
$\omega<0.1$, suggesting gapless excitations \cite{I13,I17}. On the
other hand, in the SDW state at $T=0$, the spin excitation intensity
is close to zero below $\omega\approx0.13$, similar to a spin gap
\cite{I13,I12}. However, the gap is not sharp since a sharp gap
would produce a stepwise increase in intensity at the gap energy
which is unlike the more gradual increase seen here. Figures
\ref{phase diagram}(c) and \ref{phase diagram}(d) show
$Im\chi^{RPA}(\mathbf{q},\omega)$ as a function of energy transfer
$\omega$ and momentum $\mathbf{q}$ along
$(q_{x}/\pi,q_{y}/\pi)=(h,h)$ direction at $T=0$. As we can see,
there is almost no detectable intensity below $\omega\approx0.13$,
again illustrating the opening of the spin gap. The excitations are
peaked at $\mathbf{Q}=(\pi,\pi)$ and $\omega\approx0.93$, at higher
energies, the response is seen to split and broaden due to the
dispersion of the spin waves. By tracking the peak positions in Fig.
\ref{phase diagram}(c), the spin-wave dispersion relation can be
fitted as $\omega_{q}=\sqrt{\Delta^{2}+v^{2}q^{2}}$
\cite{I11,I13,I18}, where $\Delta\approx0.93$ is an energy gap,
$v\approx9.02$ is the spin-wave velocity, and $q$ is the reduced
wave vector away from $(1,1)$ along the $(h,h)$ direction,
consistent with the experimental observations \cite{I12,I13}. The
origin of the spin gap can be understood in terms of the Fermi
surface at $x=0$ as shown in Fig. 2(a) in Ref. \cite{M4}. In the
paramagnetic state, large parts of the two hole pockets around
$\Gamma=(0,0)$ and two electron pockets around $M=(\pi,\pi)$ are
nested by momentum $(\pi,\pi)$, thus giving rise to the gapless
excitations at $T=0.27$. But at $T=0$, the SDW order will gap most
parts of the original Fermi surface, leaving only tiny ungapped
Fermi surface pockets connected by $(\pi,\pi)$ along the $\Gamma-M$
line, so in this case, for small energies, the imaginary part of the
MF spin susceptibility is close to zero, while its real part does
not fulfill the resonance condition, leading to a spin gap opening
in the RPA spin susceptibility.

The RPA spin susceptibility [Fig. \ref{Constant-Q scan}(a)] at
$x=0.06$ suggests that the excitations above $T_{c}$ are gapless,
although the intensity is very small at low energy. This may be the
reason why above $T_{c}$, Ref. \cite{I4} claims the excitations are
gapless while Ref. \cite{I10} concludes they are gapped. Below
$T_{c}$, the intensities below $\omega\approx0.06$ and above
$\omega\approx0.36$ are suppressed and the weight is transferred to
form a resonance at $\omega_{res}\approx0.14$. Since
$m(T=0.09)\approx0.129$ and $m(T=0)\approx0.131$, our results seem
to agree with Ref. \cite{I4}, which claims the resonance is produced
by suppressing low energy spectral weight, rather than Ref.
\cite{I10}, where the spectral weight is considered to be
transferred from the ordered magnetic moments. In addition, figure
\ref{Constant-Q scan}(b) shows that below $\omega_{res}$,
commensurate spin excitation prevails, in agreement with
experimental observation \cite{I10}, and it becomes incommensurate
when the energy is above $\omega_{res}$, notably between
$\omega\approx0.15$ and $\omega\approx0.18$, which we predict to be
measurable by NS experiment. The spin excitations can extend beyond
$\omega=1$, with smeared out and broadened features for
$\omega\gtrsim0.2$.

At $x=0.1$, the SDW order is completely suppressed, and SC emerges
for $T<T_{c}\approx0.06$. The excitation spectrum [Fig.
\ref{Constant-Q scan}(c)] shows that in the superconducting state at
$T=0$, a gap below $\omega\approx0.08$ develops and there is a
resonance above the gap energy peaking at $\omega_{res}\approx0.24$,
in agreement with the NS experiments on the optimally doped
Ba(Fe$_{1-x}$Co$_{x}$)$_{2}$As$_{2}$ \cite{I15,I16,I14,I17}.
Furthermore, figure \ref{Constant-Q scan}(d) shows that the spin
excitation is incommensurate at low energy
($0.18\lesssim\omega\lesssim\omega_{res}$) which still need to be
verified by experiments, then it switches to a commensurate behavior
between $\omega\approx\omega_{res}$ and $\omega\approx0.3$, and
becomes broad at higher energy, consistent with Refs.
\cite{I18,I15,I16}. In the normal state at $T=0.12$, the spectrum is
replaced by broad gapless excitations with a linear energy
dependence for $\omega<0.1$ \cite{I16,I17}. We notice a marked
similarity between the spin excitations in the normal state of the
optimally doped compound and those in the paramagnetic state of the
parent compound as observed in Ref. \cite{I17}, suggesting a common
origin of spin fluctuations in both of them.

In contrast, the spin excitations in the overdoped ($x=0.2$)
compound [Fig. \ref{Constant-Q scan}(e)] show gap-like behavior in
both the normal and superconducting states. The origin of the gap
may be due to one of the two hole pockets around $\Gamma$ vanishes
and the other one shrinks dramatically in the overdoped region
according to ARPES experiments \cite{M2} and theories \cite{M1,M4}.
Under such a case, due to the lack of interband scattering between
the hole and electron pockets, the imaginary part of the spin
susceptibility is strongly suppressed and gives rise to the
pseudogap behavior \cite{R1} which has been observed in NMR
\cite{R2} and NS \cite{I17} experiments in the electron overdoped
Ba(Fe$_{1-x}$Co$_{x}$)$_{2}$As$_{2}$. But in Ref. \cite{R1}, the
pseudogap is associated with the vanishing of one of three hole
pockets around $\Gamma$, where experimentally there is only one hole
pocket at this doping level as observed by ARPES \cite{M2}. The spin
excitations in the superconducting state at $T=0$ [Fig.
\ref{Constant-Q scan}(f)] are broader and weaker than those in the
underdoped and optimally doped compounds, suggesting the importance
of the hole pocket in enhancing the spin fluctuations.

At $x=0.3$, both the two hole pockets around $\Gamma$ disappear
\cite{M1,M4}, our calculations show that SC is completely suppressed
and the spin fluctuations are extremely small [The inset of Fig.
\ref{Constant-Q scan}(e)]. This further indicates the correlation
between the electronic band structure and magnetism, and supports
the scenario that the spin fluctuations in the underdoped regime,
which serve as a precursor to SC, originate from quasiparticle
scattering across the electron and hole pockets.

In summary, we have systematically investigated the doping
dependence of spin excitations in
Ba(Fe$_{1-x}$Co$_{x}$)$_{2}$As$_{2}$, ranging from the parent to
overdoped regime. In the parent compound, the spin excitations are
gapless in the paramagnetic state and become strongly suppressed at
low energy in the SDW state due to the opening of gaps on most parts
of the original Fermi surface. For underdoped and optimally doped
samples, the spin gaps and resonances at $(\pi,\pi)$ only occur in
the SC state. On the other hand, the spin excitations in the
overdoped compound show gap-like behavior in both the normal and SC
states due to the vanishing of one hole pocket around $\Gamma$,
leading to a ``pseudogap'' behavior at this doping level.  All the
obtained results are in qualitative agreement with experiments. The
changes in the spin dynamics at different doping levels may reflect
changes in the electronic band structure and suggest a strong
correlation between SC and magnetism.

{\it Acknowledgments} We thank D. G. Zhang, C. H. Li, J. P. Hu and
J. X. Zhu for helpful discussions. This work was supported by the
Texas Center for Superconductivity and the Robert A. Welch
Foundation under grant numbers E-1070 (Yi Gao and W.P. Su) and
E-1146 (Tao Zhou and C. S. Ting).

\end{document}